\documentclass{aastex}
\usepackage{spr-astr-addons}
\usepackage{url}

\RequirePackage{color}
\renewcommand{\vec}[1]{\mbox{\boldmath $#1$}}

\newcommand{\emaila}{wilhelm@mps.mpg.de}
\newcommand{\emailb}{bholadwivedi@gmail.com}

\newcommand*\Del{\mathrm{\Delta}}                 
\newcommand{\rmd}{ {\ \mathrm d} }
\newcommand{\uvec}[1]{\hat{\vec #1}}
\newcommand{\m}{ {\ \mathrm m} }
\newcommand{\s}{ {\ \mathrm s} }

\begin{document}

\title{An impact model of the electrostatic force: \newline
Coulomb's law re-visited}
\shorttitle{An impact model of the electrostatic force}
\shortauthors{K. Wilhelm, B.N. Dwivedi and H. Wilhelm}

\author{Klaus Wilhelm}
\affil{Max-Planck-Institut f\"ur Son\-nen\-sy\-stem\-for\-schung
(MPS), 37077 G\"ottingen, Germany\\ \emaila}
\and
\author{Bhola N. Dwivedi}
\affil{Department of Physics, Indian Institute of Technology
(Banaras Hindu University), Varanasi-221005, India \\ \emailb}
\and
\author{Horst Wilhelm}
\affil{Deceased}

\today

\vspace{1cm}

\begin{abstract}
The electrostatic force is described in this model by the action of
electric dipole distributions on charged particles. The individual
hypothetical dipoles are propagating at the speed of light in vacuum
transferring momentum and energy between charges
through interactions on a local basis. The model is constructed in analogy to
an impact model describing the gravitational forces.
\end{abstract}

\section{Introduction} 
\label{s.intro}


Electrostatic fields can be described by a subset of
Maxwell's equations:
%
\begin{equation}
\oint\vec E\,\rmd \vec s = 0 \quad {\rm and} \quad
\varepsilon_0\int\!\!\!\int_F\vec E\,\rmd \vec f
= Q_F ~ .
\label{eqn.Maxwell}
\end{equation}
$\vec E$ is the electric field strength,
d$\vec s$ an element along a closed integration path,
d$\vec f$ an element of the
closed surface~$F$ of a macroscopic volume with its normal directed outwards,
$Q_F$ the total electric charge enclosed by $F$, and
$\varepsilon_0 = 8.854...~\times~10^{-12}~{\rm F}\,\m^{-1}$
the electric constant
in vacuum. Experience has shown that electric monopoles\,--\,either as
positive or negative charges\,--\, are attached to matter, and that $Q_F$ is
zero in vacuum.

For two particles~A and B charged with $Q$ and $q$, respectively, at a
separation distance, $r$, (large compared to the sizes
of the particles) and at rest in an
inertial system, Coulomb's law yields an electrostatic force, $\vec K_{\rm E}$,
acting on $q$ of
%
\begin{equation}
\vec K_{\rm E}(\vec r) =
\frac{Q\,\uvec r}{4\,\pi\,r^2\,\varepsilon_0}~q ~ ,
\label{eqn.Coulomb}
\end{equation}
where $\uvec r$ is the unit vector of the radius
vector $\vec r$ from A to B, and $r = |\vec r|$.
If $Q$ and $q$ have opposite signs, the force is an attraction,
otherwise a repulsion. The first term on the right-hand side
%
\begin{equation}
\frac{Q\,\uvec r}{4\,\pi\,r^2\,\varepsilon_0} =
\vec E_Q(\vec r)
\label{eqn.Efield}
\end{equation}
represents the classical electric field of a charge~$Q$.
Provided $|q| \ll |Q|$, the total field will not be influenced by $q$ very
much. The electric potential, $U(r)$, of a
charge, $Q$, located at $r = 0$ is
%
\begin{equation}
U_Q(r) =
\frac{Q}{4\,\pi\,\varepsilon_0~r}
\label{eqn.Pot}
\end{equation}
for $r > 0$. The corresponding electric field can be written as
$\vec E_Q(\vec r) = - \nabla U_Q(r)$.
Eq.~(\ref{eqn.Pot}) indicates that an electric point source cannot be real
in a classical theory, and a distribution of the charge within a certain
volume must be assumed
\citep[cf. e.\,g.][and references cited therein]{Levetal,Roh97}.
The energy density of an electric field outside of charges is given by
%
\begin{equation}
w = \frac{\varepsilon_0}{2}\,\vec E^2 ~ .
\label{eqn.energy}
\end{equation}
Eqs.~(\ref{eqn.Maxwell}) to (\ref{eqn.energy}) are treated in most physics
textbooks
\citep[cf. e.\,g.][]{Hun57,Jac06}.
For a single charge, $Q$, the integration of the energy contributions
$\rmd W_Q = w\,\rmd V = 4\,\pi\,r^2\,w\,\rmd r$
from $r = \Del r$
to $r \to \infty$ yields the total field energy outside the sphere with
radius $\Del r$
%
\begin{equation}
W_Q(\Del r) = \frac{Q^2}
{8\,\pi\,\varepsilon_0}\,\int^\infty_{\Del r} \frac{\rmd r}
{r^2} = \frac{Q^2}{8\,\pi\,\varepsilon_0}\,\frac{1}{\Del r} ~ .
\label{eqn.Energy}
\end{equation}
With $Q = \pm\,|e| = \pm\,1.602\ldots \times 10^{-19}$~C,
the elementary charge, and $\Del r$ the Bohr radius
%
\begin{equation}
a_0 = \frac{\varepsilon_0\,h^2}{\pi\,m_{\rm e}\,e^2} = 52.9~{\rm pm} ~ ,
\label{eqn.Bohr}
\end{equation}
where $m_{\rm e} = 9.109\ldots \times 10^{-31}$~kg is the electron mass and
$h = 6.626\ldots \times 10^{-34}$~J\,s the Planck constant,
Eq.~(\ref{eqn.Energy}) yields
%
\begin{equation}
W_{\rm e}(a_0) = \frac{1}{2}\,\alpha^2\,m_{\rm e}\,c^2_0 ~ ,
\label{eqn.Harthalf}
\end{equation}
with $c_0 = 299\,792\,458~\m\,\s^{-1}$, the speed of light in vacuum,
and Sommerfeld's fine-structure constant
%
\begin{equation}
\alpha = \frac{e^2}{2\,\varepsilon_0\,h\,c_0} = 7.297\ldots \times 10^{-3}~.
\label{eqn.Sommerfeld1}
\end{equation}
Laboratory limits for potential
temporal drifts of the fine-structure constant $\alpha$ are given by
\citet{Haeetal,Lev04}.
They are consistent with a null result.

Choosing $\Del r = \hbar/(m_{\rm e}\,c_0) = 0.386$~pm,
the reduced Compton wavelength of
an electron, where $\hbar = h\,/\,2\,\pi$ is the
reduced Planck constant, gives
%
\begin{equation}
W_{\rm e}\left(\frac{\lambda_{\rm C}}{2\,\pi}\right) =
\frac{1}{2}\,\alpha\,m_{\rm e}\,c^2_0 ~ .
\label{eqn.Compton}
\end{equation}
With $\Del r = r_{\rm e} = \alpha^2\,a_0 = 2.82$~fm,
the classical electron radius, however,
half the rest energy of an electron will be obtained:
%
\begin{equation}
W_{\rm e}(r_{\rm e}) = \frac{m_{\rm e}\,c^2_0}{2} ~ .
\label{eqn.rest}
\end{equation}

Applying Eq.~(\ref{eqn.energy})
to a plane plate capacitor with an area, $C$,
a plate separation, $b$,
and charges $\pm\,|Q|$ on the plates, the energy stored in the field
of the capacitor turns out to be
%
\begin{equation}
W = \frac{\varepsilon_0}{2}\,\vec E^2\,C\,b =
\frac{\varepsilon_0}{2}\,\vec E^2\,V ~ .
\label{eqn.Capacitor}
\end{equation}
With a potential difference $\Del \,U = |\vec E|\,b$
and $Q = \varepsilon_0\,|\vec E|\,C$
(incrementally increased to these values), the potential energy
of the charge, $Q$, at $\Del \,U$ is
%
\begin{equation}
W = \frac{1}{2}\,Q\,\Del \,U ~ .
\label{eqn.Cenergy}
\end{equation}
The question as to where the energy is actually stored,
\citet{Hun57}
answered by showing
that both concepts implied by
Eqs.~(\ref{eqn.Capacitor}) and (\ref{eqn.Cenergy}) are equivalent.

\section{What are the problems?} 
\label{s.problem}

Maxwell's equations provide an adequate framework
for describing most electromagnetic processes.
However, as is evident from
the example of the plane plate capacitor, Maxwell's equations do neither
answer the question of the physical nature of the electrostatic field
nor that of the Coulomb force.
A detailed discussion of action at a distance
versus near-field interaction theories and hypotheses was presented by
\citet{Dru97} at the end of the nineteenth century.
Although many physicists feel that such questions are irrelevant,
because an adequate mathematical description is all that can be
expected, these problems will be considered here.

\citet{Pla09}
asked, in defence of the electromagnetic field
as energy and momentum carrier against the
idea of light quanta \citep[to become later known as photons,][]{Lew26}:
,,Wie soll man sich z.\,B. ein elektrostatisches Feld denken? F\"ur dieses
ist die Schwingungszahl $\nu = 0$, also m\"u{\ss}te die Energie des Feldes doch
wohl bestehen aus unendlich vielen Energiequanten vom Betrage {\it Null}. Ist
denn da \"uberhaupt noch eine endliche, bestimmt gerichtete Feldst\"arke
definierbar?''
(How should one imagine, e.\,g., an electrostatic field? For such a field
the frequency is $\nu = 0$, and thus the energy of the field must consist of
an infinite number of energy quanta with zero value. How is it then
possible to define a finite, unambiguously directed field strength?)

Planck obviously considered only photons as potential
carriers of the electrostatic field,
before rejecting them. Nevertheless, he acknowledged a problem in this context.
A critique of the classical field theory by
\citet{WheFey}
concluded that a theory of action at a distance,
originally proposed by
\citet{Sch03},
avoids the direct notion of field
and the action of a charged particle on itself.
However, the consequences that particle interactions occur
not only by retarded, but also by
advanced forces are not very appealing
\citep[cf.][]{Lan50}.

Near a charge~$Q$, in the well-established regions of the field,
there cannot be a ``flow'' of the electric field
with the speed~$c_0$, because the
energy density falls off with $r^{-4}$ according to Eqs.~(\ref{eqn.Efield})
and (\ref{eqn.energy}), whereas the increase
%
\begin{equation}
\Del V_r = 4\,\pi\,r^2\,c_0\,\Del t
\label{eqn.volume}
\end{equation}
of the spherical volume
covered by the flow within the time interval~$\Del t$
is only proportional to~$r^2$.

A simple model for the electrostatic force can be obtained by
introducing hypothetical electric dipole particles. A quadrupole model has
been introduced for the gravitational force by \citet{Wiletal}.
If, indeed, the gravitational forces can be described by such a model,
it appears to be unavoidable that a model
without far-reaching fields must be possible for Coulomb's law as well,
in particular, as both forces obey an $r^{-2}$ law.
The idea that these realms of physics might have a
very close connection was already expressed in 1891
by \citet{Hea92}: `` \ldots
in discussing purely electromagnetic speculations, one may be within stone's
throw of the explanation of gravitation all the time''.

The point of
view will be taken that there is no far-reaching electrostatic field,
and that the interactions have to be understood on a local basis with energy
and momentum transfer by the multipoles. In other words, we want to
propose a physical process that allows us to understand the electrostatic
equations in Sect.~\ref{s.intro}. There can, of course, be no doubt that
they are valid descriptions.

\section{A dipole model of the electrostatic force} 
\label{s.electrostatic}

\subsection{Dipole definition}
\label{ss.dipole}

Assuming a charge~$Q$ at $r = 0$
as well as a smaller one $q$ at $\vec r$,
we get from Eq.~(\ref{eqn.Pot}) the electric potential $U(r)$, and from
Eq.~(\ref{eqn.Coulomb}) with $r = ||\vec r||$ the electrostatic force.
Can there be a model description in line with the goals outlined in
Sect.~\ref{s.problem}? Far-reaching electrostatic fields can be avoided with
the help of a
model similar to the emission of photons from a radiation source
and their absorption or scattering
somewhere else\,--\,thereby transferring
energy and momentum with a speed of $c_0$ in vacuum \citep{Poi00,Ein17,Com23}.

Since, for the case at hand, an interaction with charged particles is a
requirement, electric charges have to be involved in all
likelihood. Electric monopoles, although they co-exist with matter, must be
excluded in vacuum according to
Eq.~(\ref{eqn.Maxwell}).
In principle,
it would be possible to postulate a neutral
(plasma-type) mixture of positive and negative monopoles, however,
opposite charges would separate near charged particles.
The next option would be electric dipoles that have to travel
at a speed of $c_0$,
and, consequently, cannot have mass.
It would therefore be necessary to assume that such dipoles can
exist independently of matter \citep[cf. also][]{Bon70,Jacetal,Kos08}.
For photons a zero mass follows from the Special Theory of Relativity (STR)
\citep{Ein05}
and a speed of light in vacuum constant for all
frequencies. Various experimental methods have been used to constrain
the photon mass to $m_\nu < 10^{-49}$~kg
\citep[cf.][]{GolNie}.

\begin{figure}[t]
\begin{center}
\includegraphics[width=\columnwidth]{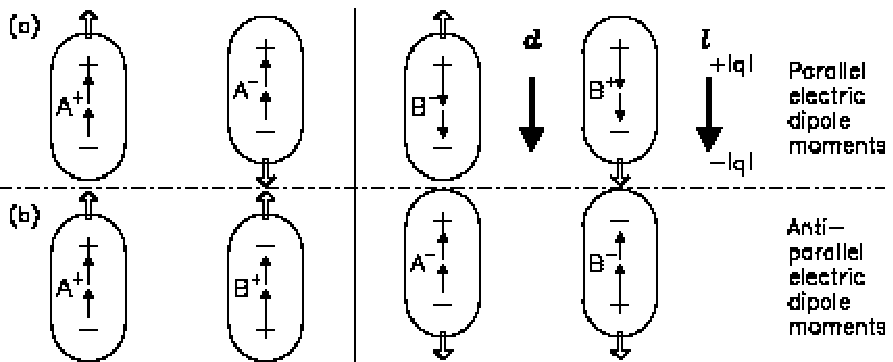}
\end{center}
\caption{Electric dipoles
with an electric dipole moment, $\vec d$, each.
They can be represented by
charges $\pm\,|q|$ and a separation vector, $\vec l$
(explicitly shown for the upper right dipole with a length, $l$ in their
rest frame). Parallel spin vectors (small arrows) are introduced for the
opposite charges resulting in no magnetic dipole moment.
There are two types of dipoles, designated by A and B, each with a
positive (A$^+$, B$^+$) and negative (A$^-$, B$^-$) spin
with respect to the velocity vectors (open arrows).
Various combinations of these bosons are
displayed that would represent: ({\rm a}) classical
electrostatic fields on larger scales and
({\rm b}) no mean electrostatic fields, but in all cases no magnetic fields.
\label{fig.Electric}}
\end{figure}

Apart from the requirement that the absolute values of the positive and
negative charges must be equal, nothing is known,
at this stage, about the values
themselves, so charges of~$\pm\,|q|$ will be assumed, which might or
might not be elementary charges $\pm\,|e|$. Such a dipole model
is sketched in
Fig.~\ref{fig.Electric}
in several configurations.
A spin $s = \hbar/2$
has been added for each charge in such a way that no combined
magnetic moment arises. The resulting total spin
is $S = \pm\,\hbar$ oriented relative to the dipole vector.
The speculation that the stability of these entities might hinge on the
electrostatic attraction of the charges, on the one
hand, and the impossibility of disposing the angular momentum should the
charges annihilate, on the other hand, was
the major reason for introducing the spin.
%
The dipole moment of the postulated particles is
%
\begin{equation}
\vec d = |q|\,\vec l
\label{eqn.EMoment}
\end{equation}
parallel or antiparallel to the velocity vector, $c_0\,\uvec n$.
This assumption is necessary in order to get attraction and repulsion
of charges depending on their mutual polarities. In Sect.~\ref{ss.density} it
will, however, be shown that the value of the dipole moment is not critical.
The dipoles have a mean kinetic energy,
$\overline{T_{\rm D} ^\infty} > 0$,
and a momentum
%
\begin{equation}
\overline{\vec p_{\rm D}} =
\frac{\overline{T_{\rm D}^\infty}}{c_0}\,\uvec n ~ ,
\label{eqn.Momentum}
\end{equation}
where $T_{\rm D} ^\infty$ represents the energy spectrum of the dipoles.
As a working hypothesis, it will first be assumed that
$T_{\rm D} \equiv \overline{T_{\rm D} ^\infty}$ is constant
remote from any charges with the same value
for all dipoles of an isotropic distribution.
The question what determines the energy of a dipole will be left open,
but one idea could, however, be
that the frequency of a longitudinal oscillation of the charges is
of importance. In Sect.~\ref{s.move}, it will be necessary to consider the
energy spectrum of the dipoles.

\subsection{Interactions with charged particles} 
\label{ss.interact}

Next it will be postulated that a particle with
charge, $Q$, and mass, $m_Q$, is emitting directed dipoles with
$\vec p_{\rm D} $. The emission
rate should be proportional to its charge, and the orientation such that a
repulsion exists between the charge and the dipoles.
The details of the process leading
to the ejection cannot be specified at this stage. The far-field aspects will
be the main interest here. The propagating dipoles
are considered to be real particles and not virtual ones as assumed in
the Quantum Field Theory.
First a description will, as much as possible, be given along the lines of
\emph{classical field concepts}, although the interpretation is later very
\emph{different}.
The conservation of energy and momentum will be taken into account,
but neither any quantum electrodynamic
aspects nor complications introduced by a particle spin. A mass~$m_Q$ of the
charge~$Q$ has explicitly been mentioned,
because the massless dipole charges are not assumed to absorb and emit
any dipoles themselves.
The conservation of momentum could hardly be fulfilled in such a process.

From energy conservation it follows that a
charge must be able to absorb dipoles as well. Otherwise it has to
be assumed that the emission is powered by a reduction of $m_Q$; an
option that will be excluded.
Therefore, the vacuum is thought
to be permeated by dipoles that are, in the absence of near charges,
randomly oriented and directed with
(almost) no interaction among each other. Note that such dipoles
have no mean interaction energy, even in the classical theory
\citep[see e.\,g.][]{Jac06}.
Their number density in space
be $\rho_{\rm E} = \Del N_{\rm E}/\Del V$, assumed to be a nearly constant
quantity, possibly slowly varying in space and time. Whether this
``background dipole radiation'' and the ``quadrupole radiation''
proposed by \citet{Wiletal} to model the gravitational attraction
are related to the dark matter and dark energy
problems is of no concern here.

A charge, $Q$, absorbs and emits dipoles at a rate
%
\begin{equation}
\frac{\Del N_Q}{\Del t}
= \kappa_{\rm E}\,\rho_{\rm E}\,|Q|
= \eta_{\rm E}\,|Q|
~ ,
\label{eqn.Emission}
\end{equation}
where $\eta_{\rm E}$ and $\kappa_{\rm E}$ are the
corresponding (electrostatic) emission and absorption
coefficients\,--\,the latter will, however, be called \emph{interaction}
coefficient, because it controls the interaction rate of a charge with a
certain spatial dipole density.
The momentum conservation can, in general, be fulfilled by isotropic
absorption and emission processes. The assumptions
as outlined will lead to a distribution of the emitted dipoles in the rest
frame of an isolated charge, $Q$, with a spatial density of
%
\begin{eqnarray}
\rho_Q(r) = \frac{\Del N_Q}{\Del V_r} =
\frac{1}{4\,\pi\,r^2\,c_0}\,\frac{\Del N_Q}{\Del t} =
\frac{\eta_{\rm E}}{c_0}\,\frac{|Q|}{4\,\pi\,r^2} ~ ,
\label{eqn.Distribution}
\end{eqnarray}
where $\Del V_r$ is given in Eq.~(\ref{eqn.volume}).
The radial emission is part of the background,
which has a larger number density
than $\rho_Q(r)$
at most distances, $r$, of interest.
Note that the emission of the
dipoles from $Q$ does not change the number density,
$\rho_{\rm E}$, in the environment of the charge,
but reverses the orientation of about \emph{half} of the dipoles affected.
Details of the interaction involving virtual dipoles to achieve this
orientation are presented in
Sect.~\ref{ss.interact} with special reference to Fig.~\ref{fig.indirect}.

The total number of dipoles will, of course, not be changed either.
For a certain $r_Q$, defined
as the charge radius of $Q$, it has to be
%
\begin{equation}
\rho_{\rm E} =
\left [\frac{\Del N_Q}
{\Del V_r} \right ]_{r_Q}
= \frac{\eta_{\rm E}}{c_0}\,\frac{|Q|}{4\,\pi\,r^2_Q}
~ ,
\label{eqn.approximation}
\end{equation}
because all dipoles of the background that come so close interact
with the charge $Q$ in some way.
At this stage, this is a formal description
awaiting further quantum electrodynamic studies in the near-field region
of charges. The quantity
$\rho_Q(r)$ for $r > r_Q$ depends on the dipole emission at the time
$t_0 = t_Q - r/c_0$
in line with the required retardation.

The same arguments apply to a charge $q \ne Q$. Since $\rho_{\rm E}$
cannot depend on either $q$ or $Q$, the quantity
%
\begin{equation}
\sigma_{\rm E}
= \frac{|q|}{4\,\pi\,r^2_q}
= \frac{|Q|}{4\,\pi\,r^2_Q}
= \frac{|e|}{4\,\pi\,r^2_{\rm E}}
\label{eqn.surface}
\end{equation}
must be independent of the charge, and can be considered as a kind of
surface charge density,
cf. ``Fl\"achenladung'' of an electron defined by \citet{Abr02},
that is the same for all charged particles.
The equation shows that $\sigma_{\rm E}$ is determined by the electron
charge radius, $r_{\rm E}$, for which estimates will be provided
in Sect.~\ref{ss.density}.
From Eqs.~(\ref{eqn.Emission}) to (\ref{eqn.surface}) it then follows that
%
\begin{equation}
\kappa_{\rm E} \, \sigma_{\rm E} = c_0 ~ .
\label{eqn.kappa}
\end{equation}
The directed dipole flux would lead to a polarization
of the vacuum with an apparent electric field~$\vec E_{\rm P}$ that can be
treated in analogy to the polarization
$\vec{P} = (\varepsilon - \varepsilon_0)\,\vec{E}$ of
matter in the classical theory \citep[cf. e.\,g.][p.~60]{Hun57}.
In the latter case for $\varepsilon \ne \varepsilon_0$,
the polarization is a consequence of the
electric field. In our case, however, the polarization is caused by the
emission characteristics of charged particles and can be interpreted as an
electric field.
Note in this context that the Lorentz
transformations do not affect the electric field component in the boost
direction.
Applying the usual polarization definitions ,
$\vec E_{\rm P}$ can be written as
%
\begin{equation}
\vec E_{\rm P}(r) =
- \frac{\vec d}
{\varepsilon_0}\,\frac{\Del N_Q}{\Del V_r}
\label{eqn.Polarization}
\end{equation}
and, with Eqs.~(\ref{eqn.Efield}) and (\ref{eqn.Distribution}),
$|\vec d| = c_0 /\eta_{\rm E}$.

In such an electric field, $\vec E$,
the dipole potential energy would be
%
\begin{equation}
T_d = -\vec d\cdot\vec E ~ .
\label{eqn.Poten}
\end{equation}
At this point, the question of the energy density of the
electric field in Eq.~(\ref{eqn.energy}) will be taken up again\,--\,the
example of the plane plate capacitor in
Sect.~\ref{s.intro}.
Eq.~(\ref{eqn.Polarization}) can be generalized to
%
\begin{equation}
\vec E =
- \frac{\vec d}
{\varepsilon_0}\,\frac{\Del N_\|}{\Del V} ~ ,
\label{eqn.general}
\end{equation}
where $\Del N_\|/\Del V$ denotes the density of the aligned dipoles
without reference to a central charge. For completeness it should be
mentioned that two or more aligned dipole distributions have to be added
taking their vector characteristics into account, in the same way as
classical electric fields. This also applies, when the distributions of
many charges have to be evaluated. The aim of the dipole model is to provide
a physical process describing the interaction of charged particles and not
to disprove Eqs.~(\ref{eqn.Maxwell}) to ({\ref{eqn.Efield}).
According to Eq.~(\ref{eqn.Poten}), this assembly of
dipoles has (in the electric field $\vec E$)
the potential energy density
%
\begin{equation}
w_{\rm P} =
-\vec d\cdot\vec E\,\frac{\Del N_\|}
{\Del V}
~ .
\label{eqn.potenergy}
\end{equation}
By turning the orientation of half of them around, the electric field would
vanish. The energy density of the electric field thus is with
Eq.~(\ref{eqn.general})
%
\begin{equation}
w = \frac{w_{\rm P}}{2} = \frac{\varepsilon_0}{2}\,\vec E^2 ~ .
\label{eqn.EnergyDensity}
\end{equation}
The energy density of the field is just a consequence of the
dipole distributions and orientations. Outside of the plane plate capacitor
discussed in Sect.~\ref{s.intro},
the orientations are such
that they compensate each other, and it is $w = 0$.

This description of the relationship between the configuration of the dipole
distribution and the electric field
suffers from the fact that both the classical electric field and its
interaction with electric charges and dipoles do not exist in the new concept.
Consequently, the classical equations of charged-particle physics as
summarized, for instance, by \citet{Roh97} cannot be applied, and
alternative solutions have to be conceived.

\subsection{Virtual dipoles} 
\label{ss.virtual}

\begin{figure}[t]
\begin{center}
\includegraphics[width=\columnwidth]{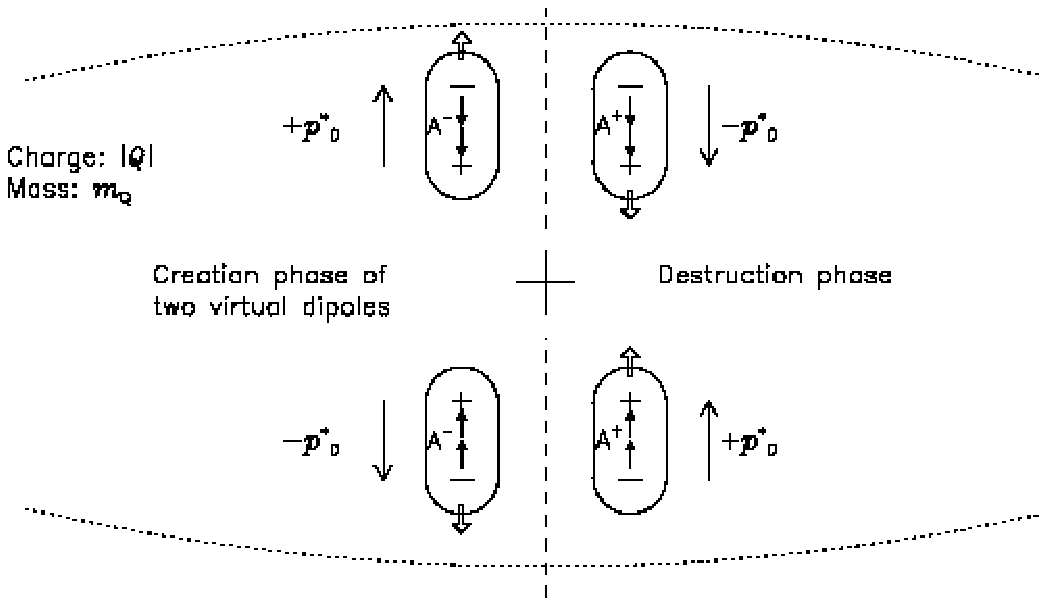}
\end{center}
\caption{Conceptional presentation of the creation
and destruction phases
of virtual dipole pairs by a charge, and the corresponding momentum vectors
of the virtual dipoles.
\label{fig.Creation}}
\end{figure}
As a first step, a formal way of achieving the required momentum and
energy transfers by discrete interactions will be described: the idea is
based on virtual dipoles
in analogy to other virtual particles \citep[cf. e.\,g.][]{Wei34,Yuk35,NimSta}.
The symmetric creation and destruction of virtual dipoles is sketched in
Fig.~\ref{fig.Creation}.
The momentum balance is shown for the emission phase on
the left and the absorption phase on the right.

Virtual dipoles with energies of $T^*_{\rm D} \ll m_Q\,c^2_0$ supplied by
a central body with charge~$Q$ and mass~$m_Q$ will have a certain lifetime
$\Del t_{\rm E}$
and interact with ``real'' dipoles. This concept implies that electric
point charges do not exist. Any charge has an interaction radius (cf.
Eq.~\ref{eqn.Pot} for the classical theory).
In the literature, there are many different derivations of an energy-time
relation \citep[cf.][]{ManTam,AhaBoh,Hil98}.
Considerations of the spread of the frequencies of a limited wave-packet
led \citet{Boh49} to an approximation for the indeterminacy of the energy
that can be re-written as
%
\begin{equation}
T^*_{\rm D} \approx h/\Del t_{\rm E}
\label{eqn.lifetime}
\end{equation}
with $h = 6.626\ldots \times 10^{-34}$~J\,s, the Planck constant. For
propagating dipoles, the equation
%
\begin{eqnarray}
l_{\rm E} = c_0\,\Del t_{\rm E} = \frac{h}{p_{\rm D} }
= \frac{h\,c_0}{T_{\rm D}}
\label{eqn.Heisenberg}
\end{eqnarray}
is equivalent to the photon energy relation
$E_\nu = h\,\nu = h\,c_0/\lambda$, where $\Del t_{\rm E}$ now corresponds to
the period and $\lambda$ to $l_{\rm E}$,
which can be considered as the de Broglie wavelength of the
hypothetical dipoles and their interaction length.
The relationship between the interaction length, defined
in Eq.~(\ref{eqn.Heisenberg}), and the charge radius in
Eq.~(\ref{eqn.surface}) could not be elucidated in this conceptual study.
Since there is, however, experimental evidence that virtual \emph{photons}
(identified as evanescent electromagnetic modes) behave non-locally
\citep{LowMen,StaNim}, the virtual dipoles might also
behave non-locally and the absorption of a dipole can occur momentarily
by the annihilation with an appropriate virtual particle.
Eq.~(\ref{eqn.Heisenberg}) would thus not be relevant in this context.
%
\begin{figure}[t]
\begin{center}
\includegraphics[width=\columnwidth]{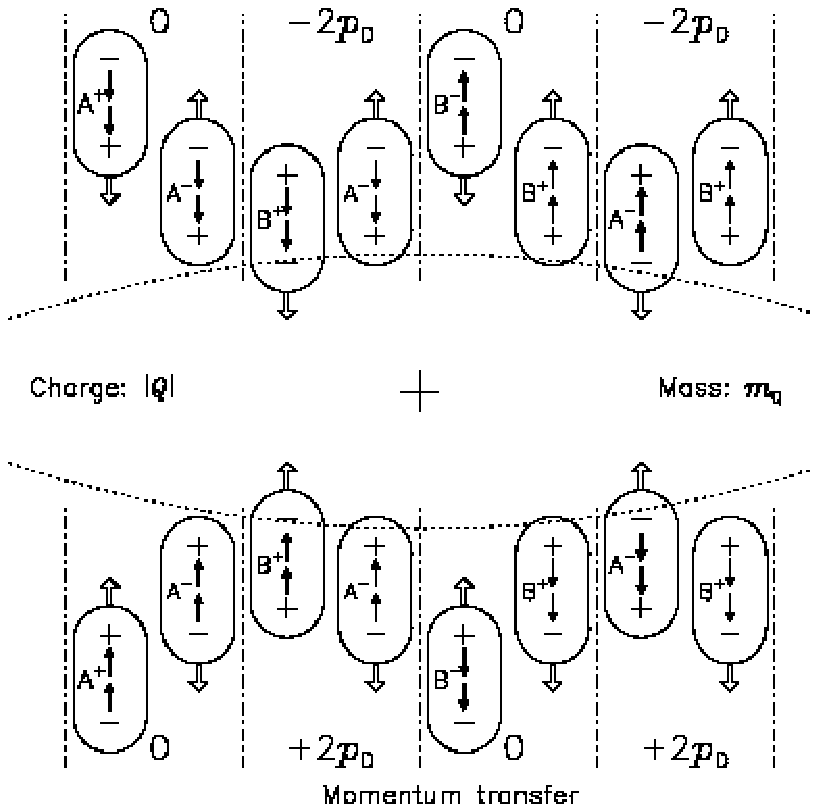}
\end{center}
\caption{Virtual dipoles of a charge $+|Q|$ (with mass $m_Q$)
interact with ``real'' dipoles, A$^+$ and A$^-$ arriving with a
momentum~$-\,p_{\rm D}$, each.
On the left, a dipole~A$^+$ annihilates in the lower
dashed-dotted box with virtual dipole~A$^+$ in the destruction phase
(cf. Fig.~\ref{fig.Creation}) and
liberates dipole~B$^+$. \emph{No momentum
will be transferred to the central charge} with $p_{\rm D}  = p^*_{\rm D} $.
The other type of
interaction\,--\,called direct interaction,
in contrast to the indirect one on the left\,--\,also
requires two virtual dipoles, one of them annihilates in the creation phase
(see again Fig.~\ref{fig.Creation}) with
dipole~A$^-$ (in the upper box with dashed-dotted boundaries),
the other one is liberated by the excess energy of the annihilation.
\emph{The central charge received a momentum of}
$ - (p_{\rm D}  + p^*_{\rm D} )$. No spin reversal has been assumed
in both cases.
\label{fig.indirect}}
\end{figure}
%
\begin{figure}[t]
\begin{center}
\includegraphics[width=\columnwidth]{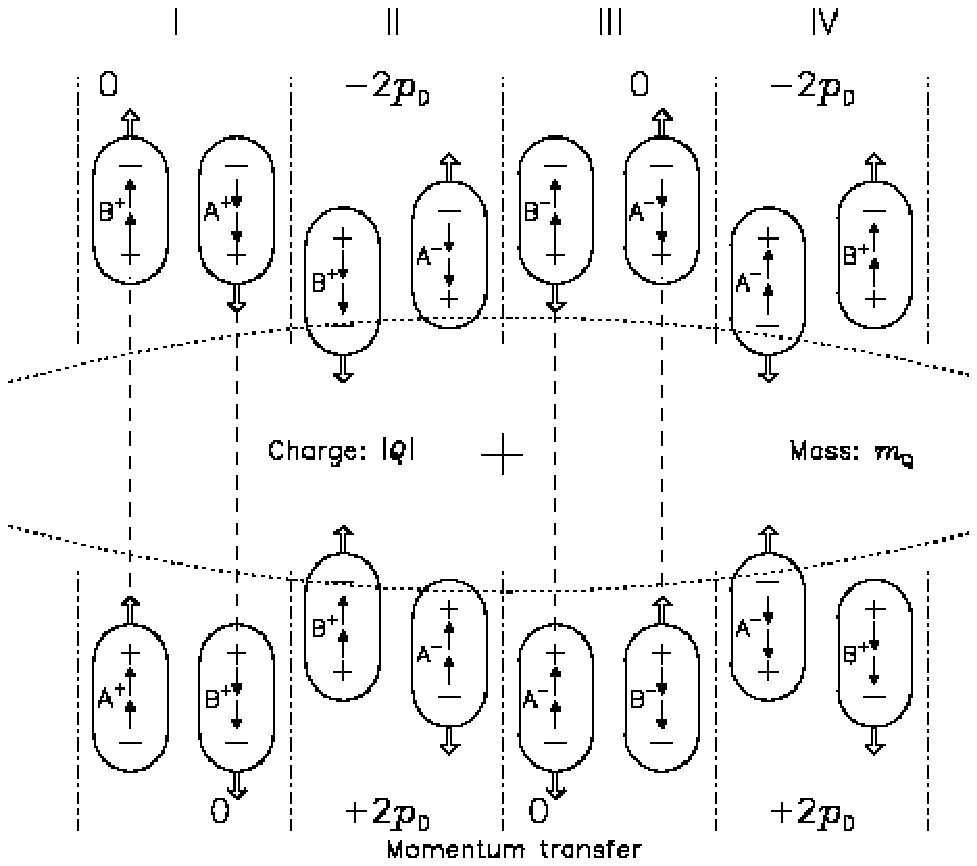}
\end{center}
\caption{Electric dipoles are approaching and leaving a charge $|Q|$
with nearly radial orientations of their dipole moments.
The central charge faces equal polarities of the outgoing
dipoles. The interaction of the charge with the dipole distribution in an
initially chaotic environment generates a configuration with
spherical symmetry. Only two directions are
shown with groups of incoming and outgoing dipoles in four vertical columns.
The interaction region is indicated by dotted lines.
(Note that the dipole trajectories are actually converging or diverging.
Together they act as an equivalent of a classical inhomogeneous electric
field.) By employing the virtual dipole interaction processes shown in
Fig.~\ref{fig.indirect} the dipoles~A$^+$ and B$^-$ arriving from above and
below in columns~I and III are ``transmitted'' without transferring
momentum to the central charge, but experience a reversal of the dipole
moment during an indirect interaction. The dipoles~A$^-$ and B$^+$ in
columns~II and IV, however, are ``reflected'' with momentum transfers as
indicated in direct interactions.
A downward momentum transfer is reckoned negative.
There is no net momentum transfer to the charge in a symmetric configuration.
\label{fig.Dipoles}}
\end{figure}

During a \emph{direct} interaction, the dipole~A$^-$
(in
Fig.~\ref{fig.indirect}
on the right side) annihilates together
with an \emph{identical} virtual dipole with an opposite velocity
vector. This postulate is motivated by the fact that it provides the
easiest way to eliminate the charges and yield
$P = -\,p_{\rm D}  + p^*_{\rm D}  = 0$ (where $p^*_{\rm D} $ is the momentum
of the virtual dipole) as well as $S = 0$ for charges at rest in a system
with an isotropic distribution of $\vec p_{\rm G}$
(cf. Sect.~\ref{ss.dipole}).
The momentum balance is neutral and the
excess energy, $T_{\rm D} $, is used to liberate a second virtual dipole~B$^+$,
which has the required orientation.
The charge had emitted two virtual dipoles
with a momentum of $+\,p^*_{\rm D} $, each, and a total momentum of
$- (p_{\rm D}  + p^*_{\rm D} )$ was transferred to $|Q|$.
The process can be described as a reflection of a dipole
together with a reversal of the dipole momentum.
The number of these direct interactions will be denoted by
$\Del \hat{N}_Q$.
The dipole of type~A$^+$ (on the left side) can
exchange its momentum in an \emph{indirect} interaction
only on the far side of the charge with an \emph{identical} virtual dipole
during its absorption (or destruction) phase (cf.
Fig.~\ref{fig.Creation}).
The excess energy of $T_{\rm D} $
is supplied to liberate a second virtual dipole~B$^+$.
The momentum transfer to the charge $+\,|Q|$ is zero.
This process just corresponds to a
double charge exchange. Designating the number of interactions of the
indirect type with
$\Del \tilde{N}_Q$, it is
%
\begin{equation}
\frac{\Del N_Q}{\Del t} =
\frac{\Del \hat{N}_Q +
\Del \tilde{N}_Q}{\Del t} ~ ,
\label{eqn.Distrib2}
\end{equation}
with $\Del \tilde{N}_Q = \Del \hat{N}_Q =
\Del N_Q/2$. Unless direct and indirect interactions
are explicitly specified, both types are meant by the term ``interaction''.

In Sect.~\ref{s.move}, moving charged bodies will be
considered. Since STR requires that a uniform motion is maintained without
external forces, it will follow
that the absorbed and emitted dipoles must not only have balanced
momentum and energy budgets in the rest system of the body, but also in other
inertial systems.

The virtual dipole emission rate has to be
%
\begin{equation}
\frac{\Del N^*_Q}{\Del t} =
2\,\frac{\Del N_Q}{\Del t} ~ ,
\label{eqn.virtualrate}
\end{equation}
i.e. the virtual dipole emission rate equals the sum of the real
absorption and emission rates.
The interaction model described results in a mean momentum
transfer per interaction of $p_{\rm D}$
\emph{without involving a macroscopic electrostatic field}.

A macroscopic dipole encounters in the
inhomogeneous dipole distribution of a charge more or less direct or indirect
interactions depending on its orientation.

\subsection{Coulomb's law} 
\label{ss.law}

As a second step, a quantitative evaluation will be considered
of the force acting on a test particle with
charge, $q$, at a distance, $r$, from another particle with charge
$Q$ by the absorption of dipoles not only from the background,
but also from the distribution emitted from $Q$ according to
Eq.~(\ref{eqn.Distribution}) under the assumption of a \emph{constant}
interaction coefficient, $\kappa_{\rm E}$.
Both charges are initially at rest in an inertial system,
and with large mass-to-charge ratios any accelerations can be neglected.
The rate of interchanges between these point sources then is
%
\begin{eqnarray}
\frac{\Del N_{Q,q}}{\Del t} =
\frac{\eta_{\rm E}\,\kappa_{\rm E}}{c_0}\,\frac{|Q|\,|q|}{4\,\pi\,r^2} =
\frac{\Del N_{q,Q}}{\Del t} ~ ,
\label{eqn.reciprocal}
\end{eqnarray}
which confirms the reciprocal relationship between $q$ and
$Q$ (after due consideration of the retardation).
It should be noted\,--\,for later reference\,--\,that
a relative motion between $q$ and $Q$ will require a relativistic treatment.
It is important to realize that all interchange events between pairs of
charged particles are either direct or indirect depending on their
polarities and transfer a momentum of $2\,p_{\rm G}$ or zero.

In order to get some more insight into the details of the process,
a positive test charge, $+\,|q|$, will be positioned below $+\,|Q|$ in
Fig.~\ref{fig.Dipoles}. Most of the dipoles arriving from below will
still be part of the background, because its density~$\rho_{\rm E}$ is much
higher than $\rho_{\rm Q}(r)$ for $r \gg r_{\rm Q}$. Assuming that \emph{one}
of them, for instance A$^+$ in column~I, has been reversed by $+\,|q|$, this
will lead to an additional direct interaction and a transfer of
$+\,2\,p_{\rm G}$.

The resulting
momentum transfer thus becomes positive leading to a repulsion, because
the conditions on the opposite side have not changed.
The momentum transfer rate from charge~ $+\,|q|$ to $+\,|Q|$ thus is with
Eq.~(\ref{eqn.reciprocal})
%
\begin{eqnarray}
\left|\frac{\Del P_{\rm E}}{\Del t}\right| =
2\,p_{\rm D}\frac{\Del N_{Q,q}}{\Del t} ~.
\label{eqn.exchange}
\end{eqnarray}
Conversely, for charges with opposite signs, dipoles emitted from $-\,|q|$
will transfer less momentum to $|Q|$, because more indirect interactions
occur. With a reversal of dipole~B$^+$ in column~II, for instance,
the balance gets negative with the consequence of an attraction, since there
is no compensation for the direct background interactions from the opposite
direction. Eq.~(\ref{eqn.exchange}) shows that the choice of the dipole
moment is of no importance in this context. With
%
\begin{equation}
2\,p_{\rm D}\,\frac{\eta_{\rm E}\,\kappa_{\rm E}}{c_0} =
\frac{1}{\varepsilon_0} ~ .
\label{eqn.eta}
\end{equation}
and taking into account the polarities, Eq.~(\ref{eqn.exchange}) becomes
%
\begin{equation}
K_{\rm E} = \frac{\Del P_{\rm E}}{\Del t} =
\frac{Q\,q}
{4\,\pi\,r^2\,\varepsilon_0} ~ ,
\label{eqn.el_force}
\end{equation}
i.e. Coulomb's law.

\begin{figure}
\begin{center}
\includegraphics[width=\columnwidth]{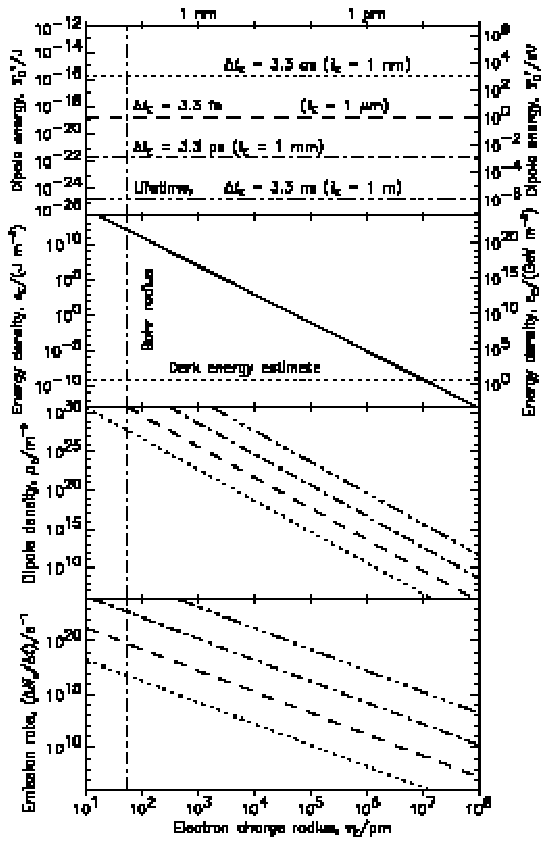}
\end{center}
\caption{Energies of the virtual dipoles according to Eq.~(\ref{eqn.lifetime})
are shown as broken lines in the upper panel
with the lifetimes, $\Del t_{\rm E}$, as parameter
(in parentheses, the interaction lengths, $l_{\rm E}$,
cf. Eq.~\ref{eqn.Heisenberg}).
In the lower panels,
the energy and number densities in space are plotted as well as
the emission rates from an elementary charge as functions of the charge
radius, $r_{\rm E}$, of a charge $e$ for different
$\Del t_{\rm E}$ values.
The Bohr radius, $a_0$, and an estimate of the cosmic dark energy
\citep[cf.][]{BecMac}
are indicated.
The charge radius consistent with the latter
value is shown as vertical dotted line.
\label{fig.Matter}}
\end{figure}

The electrostatic force between charged particles has thus been described
by the transfer of momentum through electrostatic dipoles
moving with $c_0$ and aligned along their propagation direction.
The description illustrates a physical mechanism for such a force.
In this context, it might be appropriate to note that Coulomb's interaction
energy of two electrons follows from quantum electrodynamics if only the
\emph{longitudinal} electromagnetic waves are taken into account
\citep{Fer32,BetFer}, \citep[cf.][]{Dir51a}.

\subsection{Dipole energy density} 
\label{ss.density}

Important questions obviously are related to the energy, $T_{\rm D}$, of the
dipoles
and, even more, to their energy density in space. Eqs.~(\ref{eqn.Momentum})
and (\ref{eqn.approximation}) to (\ref{eqn.kappa}) together with
Eq.~(\ref{eqn.eta}) allow the energy density to be expressed by
%
\begin{equation}
\epsilon_{\rm E} = T_{\rm D}\,\rho_{\rm E} =
\frac{\sigma^2_{\rm E}}{2\,\varepsilon_0} ~ .
\label{eqn.energydensity}
\end{equation}
This quantity is independent of both the dipole moment and energy.
It takes into account all
dipoles (whether their distribution is chaotic or not). Should the energy
density vary in space and\,/\,or time, the surface charge density,
$\sigma_{\rm E}$, must vary as well. However, as long as $\kappa_{\rm E}$ and
$\eta_{\rm E}$ obey
Eqs.~(\ref{eqn.kappa}) and (\ref{eqn.eta}),
Eq.~(\ref{eqn.exchange}) would be equivalent to Coulomb's law.

The electric dipole moment is related to the other dipole characteristics
discussed so far, only if the definition of an electrostatic
field according to Eq.~(\ref{eqn.Polarization}) is upheld.
Without that requirement, the value of the dipole moment is of no importance
and only the polarity remains a salient feature.
The other relations are shown in
Fig.~\ref{fig.Matter}
as function of the charge radius, $r_{\rm E}$, of an
electron with the lifetime, $\Del t_{\rm E}$, as parameter.
If $r_{\rm E} = a_0$, the Bohr radius,
very high energy densities result and, on the other hand, relatively large
$r_{\rm E}$ are required for energy densities in the range of the cosmic dark
energy estimates. For the time being, the latter values must be considered
as the most likely ones, and
it remains to be seen how the charge radius and the interaction length
can be further
constraint, and whether these relations are consistent with an explicit
interaction model.


\section{Dipoles and moving particles} 
\label{s.move}

Up to now, the interactions of isolated electric charges with dipoles
have been treated in an inertial system, in which the
charges were assumed to be at rest, but it is, of course,
required by STR that any uniformly moving charged particle experiences no
deceleration by the background dipole distribution. Since charges are in
general not at rest in a single inertial system, such considerations are of
great importance.
\citet{Dir51b}
argued that an {\ae}ther concept could be consistent with STR provided
quantum fluctuations of the {\ae}ther would be taken into account
\citep[cf. also][]{Vig95}.
In the meantime, experimental results have indicated that virtual particles
show a non-local behaviour \citep{LowMen,StaNim,NimSta}.
It, therefore, does not seem to be hopeless to show that charged particles at
rest in different inertial systems would be affected in
the same way by isotropic dipole distributions. This has, in fact, been shown
for moving masses relative to a quadrupole distribution \citep{Wiletal} and
can be directly applied here. The calculations for electric dipoles will give
$\Del m = 0$, because the gravitational energy reduction parameter~$Y$ is not
relevant in the electrostatic case.

\section{Discussion} 
\label{s.discuss}

Some consequences resulting from the concepts developed are discussed in the
following sections.

\subsection{Outlook} 
\label{ss.outlook}

There are several topics that, in principle, should
have been treated in this context,
but would either require detailed calculations outside the scope of this
conceptional presentation,
or further far-reaching assumptions. Within the first category fall:
(a)~forces between parallel conductors and other magnetostatic configurations,
considering that many magnetic fields can be transformed into
pure electric fields by a change of the inertial system
\citep[discussed by][]{Dwietal};
(b)~the Aharonov--Bohm effect \citep{AhaBoh59}.

In the second category should be mentioned:
(a)~the Casimir effect \citep{Cas49};
(b)~the dipole and quadrupole energy spectra;
(c)~the interaction of quadrupoles and dipoles;
(d)~the contributions of $\epsilon_{\rm E}$ and
$\epsilon_{\rm G}$ to the ``vacuum energy'' density.

\section{Conclusion} 
\label{s.concl}

The electrostatic attraction and repulsion could be described
by suitable dipole distributions tailored to emulate Coulomb's law.
This concept allows a formulation of the
far-reaching electrostatic force between charged  particles as local
interactions of postulated electric dipoles
travelling with the speed of light in vacuum.


\begin{acknowledgements}
We thank
Eckart Marsch,
Harry Kohl,
Luca Teriaca,
Werner Curdt and
Bernd Inhester
for many discussions on these topics.
Their critical comments have been
very helpful in formulating our ideas.
We also want to acknowledge constructive comments by three anonymous reviewers.
This research has made extensive use of
the Astrophysics Data System (ADS).
\end{acknowledgements}

\end{document}